\documentclass[a4paper]{jpconf}
\usepackage{graphicx}
\usepackage{amssymb}
\begin{document}
\title{Spectroscopic analysis of DA white dwarfs from the McCook \& Sion catalog\footnote{Based on observations made with ESO Telescopes at the La Silla or Paranal Observatories under program ID 078.D-0824(A) and with the Las Campanas 2.5 m Ir\'en\'ee du Pont telescope.}}

\author{A Gianninas$^1$, P Bergeron$^1$ and M T Ruiz$^2$}

\address{$^1$ D\'epartement de Physique, Universit\'e de Montr\'eal, C.P. 6128, 
Succ.~Centre-Ville, Montr\'eal, Qu\'ebec H3C 3J7, Canada}
\address{$^2$ Departamento de Astronom\'\i a, Universidad de Chile, Casilla 36-D, Santiago, Chile}

\ead{gianninas@astro.umontreal.ca,bergeron@astro.umontreal.ca,mtruiz@das.uchile.cl}

\begin{abstract}
For some years now, we have been gathering optical spectra of DA white
dwarfs in an effort to study and define the empirical ZZ Ceti
instability strip. However, we have recently expanded this survey to
include all the DA white dwarfs in the McCook \& Sion catalog down to
a limiting visual magnitude of $V$=17.5. We present here a
spectroscopic analysis of over 1000 DA white dwarfs from this ongoing
survey. We have several specific areas of interest most notably the
hot DAO white dwarfs, the ZZ Ceti instability strip, and the DA+dM binary
systems.  Furthermore, we present a comparison of the ensemble
properties of our sample with those of other large surveys of DA white
dwarfs, paying particular attention to the distribution of mass as a
function of effective temperature.
\end{abstract}

\section{Introduction}

Although the Sloan Digital Sky Survey (SDSS) has unearthed thousands
of new white dwarfs, it is our belief that there is a significantly
brighter sample of white dwarfs whose scientific potential has never
truly been exploited. With this in mind, we have undertaken a
systematic survey of DA white dwarfs based, in large part, on the last
published version of the catalog of spectroscopically identified white
dwarfs of McCook \& Sion (1999). We will
first describe our survey and how it compares to other large surveys
of DA white dwarfs and examine some of the ensemble properties of our
current sample. We will then take a closer look at a few choice
subsamples of objects. In particular, we will look at the DAO white
dwarfs, the ZZ Ceti instability strip, and the DA+dM binaries. Finally, we
will look at one specific object, CBS 229, which we believe to be a
binary system with a magnetic component.

\section{Survey Overview}

Over the last few years, we have been obtaining optical spectra for DA
white dwarfs near the ZZ Ceti instability in an effort to constrain
its empirical boundaries. More recently, we expanded this
observational survey to include all the DA white dwarfs from the
McCook \& Sion catalog down to a limiting visual magnitude of
$V$=17.5. Many of these stars have never been analyzed with modern CCD
spectroscopy and the only information available is a spectral
classification, which is often erroneous. The bulk of this project has
been conducted using Steward Observatory's 2.3 m telescope at Kitt
Peak. However, we were also able to obtain time on the ESO 3.6 m
telescope at La Silla (Chile) as well as Carnegie Observatories' 2.5 m
du Pont telescope at Las Campanas (Chile), allowing us to extend our
survey into the southern hemisphere.

\begin{figure}[h]
\begin{center}
\includegraphics[bb=50 200 525 650,width=9.5cm]{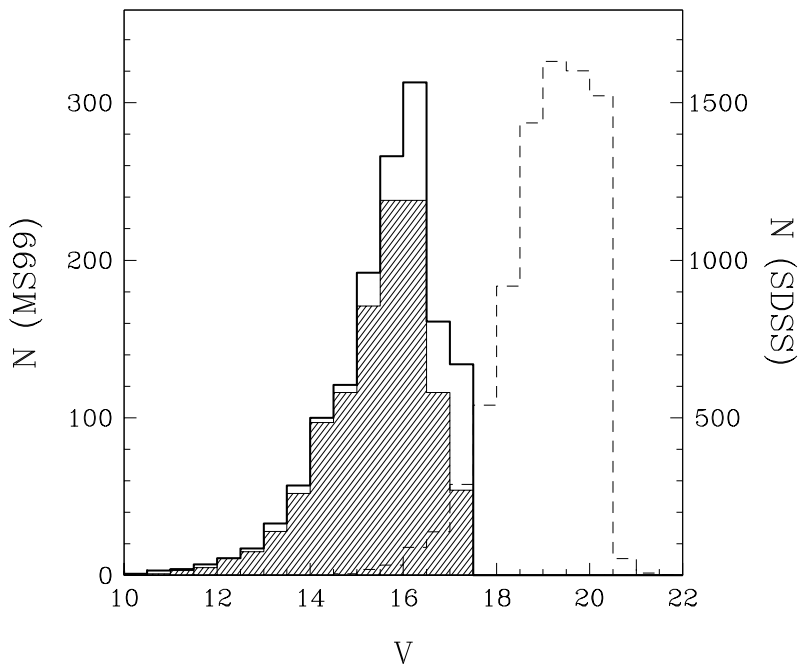}
  \caption{\label{histo}Distribution of visual magnitudes, $V$, for
    our sample selected from McCook \& Sion (1999; {\it bold
      histogram}) and for the white dwarfs we have observed to date
    ({\it hatched histogram}). In comparison, the distribution for the
    SDSS sample as of Data Release 4 (Eisenstein et al. 2006) is also
    shown ({\it dashed histogram}). Note that the scale is different
    for the McCook \& Sion sample ({\it left}) and the SDSS sample
    ({\it right}).}
\end{center}
\end{figure}

\begin{figure}[t]
\begin{center}
\includegraphics[bb=75 150 575 650,width=8cm,angle=270]{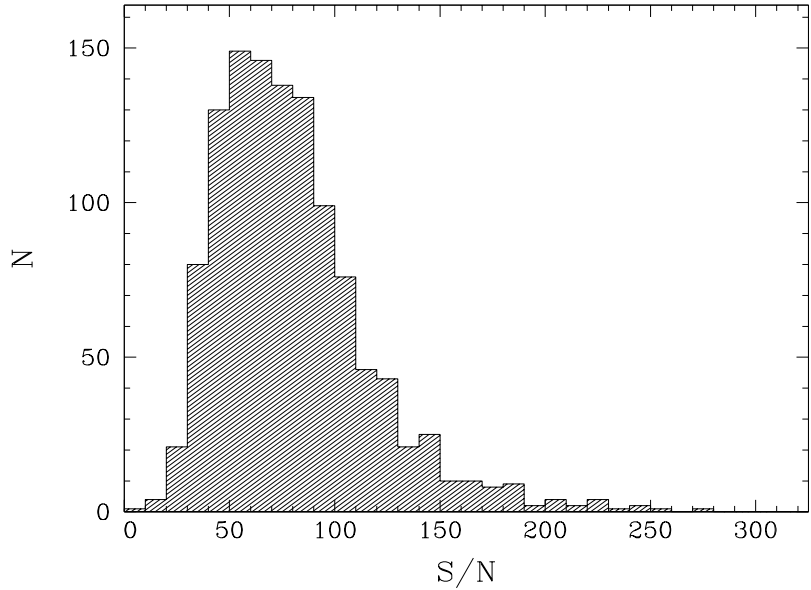}
\caption{\label{SN} Distribution of S/N for the sample of DA white
  dwarfs for which we have obtained optical spectra. The majority of
  the spectra have a S/N of 50 or greater.}
\end{center}
\end{figure}

How does our work compare to other large surveys of DA white dwarfs?
The SPY project (Koester et al. 2001), which obtained high resolution
spectra for several hundred white dwarfs, had a limiting magnitude of
$V$= 16.5 and was conducted in the southern hemisphere at the
VLT. There is actually quite a large overlap with our own sample of
stars. Indeed, close to 80\% of the stars surveyed in SPY are included
in this work. In contrast, we have only a very small overlap with the
SDSS (see Figure \ref{histo}) owing to the fact that the majority of
their objects are quite faint due to the nature of the SDSS
itself. The faintness of the objects in the SDSS also limits the
quality of their spectra since a single exposure time is set for all
the objects observed on a given plate. Therefore, the dimmer an object,
the lower the signal-to-noise ratio (S/N) of the spectrum. This makes
for an extremely inhomogeneous sample of data as far as S/N is
concerned. In contrast, we observe one star at a time and adjust our
exposure times to obtain spectra with a minimum S/N of approximately
50 (see Fig.~\ref{SN}). Thus our sample, although not as large as the
SDSS, comprises data of much higher quality overall. This is key as we
have shown in Gianninas, Bergeron \& Fontaine (2005) the importance of
high S/N for measuring the atmospheric parameters of DA white dwarfs.

\section{Mass distribution}

\begin{figure}[h]
\begin{center}
\includegraphics[bb=100 75 525 750,angle=270,width=13cm]{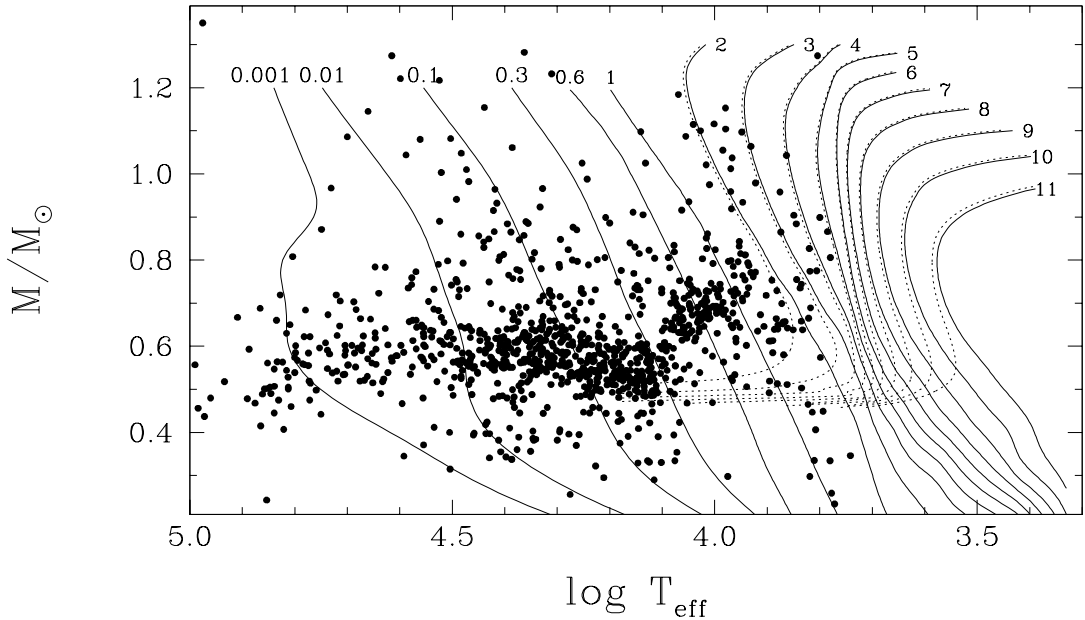}
\caption{\label{mass} Mass distribution as a function of effective
  temperature for our entire sample. The solid lines represent
  isochrones which take into account only the white dwarf cooling time
  whereas the dotted lines include also the main sequence
  lifetime. Each isochrone is labeled by its age in Gyr.}
\end{center}
\end{figure}

We show in Figure \ref{mass} the mass distribution of our sample as a
function of effective temperature. The atmospheric parameters, $T_{\rm
eff}$ and $\log g$, are determined using the spectroscopic technique
described in Bergeron, Saffer, \& Liebert (1992; see also Liebert,
Bergeron \& Holberg 2005) and the masses are derived from evolutionary
models with carbon/oxygen cores and thick hydrogen layers (see
references in Liebert et al.~2005). Our mass distribution shows the
usual increase in mass at lower temperatures as seen in the PG sample
(Liebert et al.~2005), for example. This phenomenon is still not
understood although many possible solutions have been proposed over
the years (Bergeron, Gianninas \& Boudreault 2007). We also notice a
certain number of white dwarfs with masses less than $\sim$0.45
$M_{\odot}$. This population of low mass white dwarfs is necessarily
the product of binary evolution as a progenitor with the appropriate
mass could not yet have evolved to the white dwarf stage. As such,
these objects are important as they represent a separate evolutionary
channel for white dwarf stars.

\begin{figure}[h]
\begin{center}
\includegraphics[bb=150 75 525 700,angle=270,width=13cm]{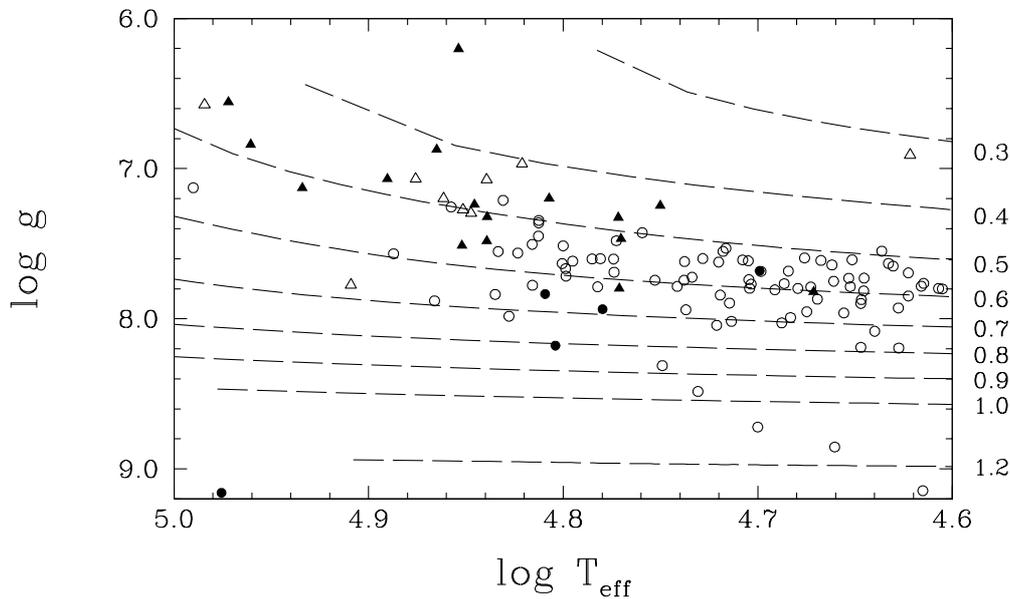}
\caption{\label{hot} Section of the $\log g$ vs $\log T_{\rm eff}$
  diagram showing the hot end of the DA white dwarf cooling
  sequence. Circles represent DA white dwarfs and triangles correspond
  to the DAO stars. The filled symbols indicate those stars that
  exhibit the Balmer line problem. The dashed lines are white dwarf
  cooling tracks with thick hydrogen layers and masses (in $M_{\odot}$)
  indicated to the right of the figure.}
\end{center}
\end{figure}

\section{DAO white dwarfs}

DAO stars are hydrogen atmosphere white dwarfs whose optical spectra
also contain lines of ionized helium, usually He {\sc ii}
$\lambda$4686. The mechanism which maintains the helium in the
atmosphere is still unclear, although a weak stellar wind has already
been proposed. Figure \ref{hot} shows the location of the DAO stars
from our sample in the $T_{\rm eff}$-$\log g$ plane along with the
regular DA white dwarfs that populate that same region of the
diagram. First, we notice that the sequence of DAO stars seems to be
best reproduced by the 0.5~$M_{\odot}$ cooling track in contrast with
the DA stars that follow the 0.6~$M_{\odot}$ cooling track, which is
consistent with the accepted mean mass for white dwarfs. This can be
explained if we assume that these DAO stars are the product of
post-EHB evolution whereby the progenitor was not massive enough to
climb back up the asymptotic giant branch (AGB) and evolved to the
white dwarf stage directly from the extreme horizontal branch
(EHB). We note, however, that this scenario does not apply to all DAO
white dwarfs. As shown in Napiwotzki (1999), there is a sequence of
DAO stars that are consistent with normal post-AGB evolution like the
majority of white dwarfs.

We also notice in Figure \ref{hot} that most of the DAO stars exhibit
the Balmer line problem as first reported by Napiwotzki (1992). The
problem manifests itself as an inability to obtain consistent values
of $T_{\rm eff}$ and $\log g$ from the spectroscopic fitting technique
for the individual Balmer lines. Werner (1999) eventually showed that
the problem could be solved by including C, N, and O in the models
with proper Stark broadening of the metallic lines. We hope to include
this solution within the next generation of our models in order to
properly analyze these DAO white dwarfs as well as the other DA stars
which exhibit the same phenomenon.

\section{The ZZ Ceti instability strip}

\begin{figure}
\begin{center}
\includegraphics[bb=100 100 525 675,angle=270,width=14cm]{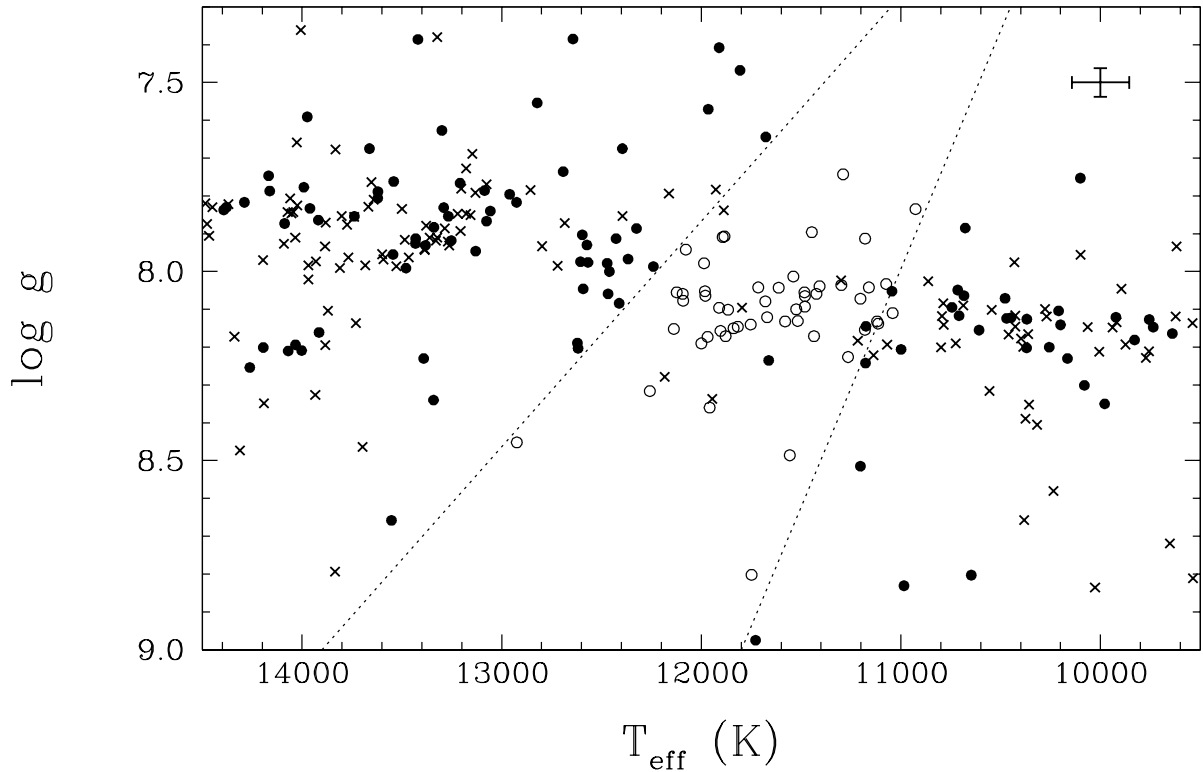}
\caption{\label{ZZ} Section of the $\log g$ vs $T_{\rm eff}$ diagram
  showing the ZZ Ceti instability strip. Open circles correspond to ZZ
  Ceti stars while filled circles represent stars that are
  photometrically constant. The $\times$ symbols are stars from our
  survey which have never been observed for variability. The dashed
  lines correspond to the empirical boundaries of the instability
  strip and the typical error bars in this region of the $T_{\rm
    eff}$-$\log g$ plane are shown in the upper right corner. }
\end{center}
\end{figure}

Figure \ref{ZZ} shows our most up to date vision of the ZZ instability
strip. In particular, we have re-observed several of the variables in
the strip and with the exception of one pulsator, we have spectra with
S/N $\gtrsim$ 70 for all our ZZ Ceti stars. However, one also notices
the presence of a photometrically constant star in the middle of the
instability strip, HS 1612+5528. This object had been reported as NOV
(not observed to vary) by Voss et al.~(2006). We obtained our own
spectrum of this star and confirmed its position within the
instability strip. Consequently, we decided to conduct our own
observations to determine whether this star was variable or not. We
observed HS~1612+5528 at the Observatoire du mont M\'egantic using the
Montr\'eal three-channel photometer LAPOUNE for several hours on the
night of 2006 July 20 and we detected no variations down to a limit of
0.2\%. However, there are known ZZ Ceti stars that pulsate with
amplitudes as low as 0.05\% so we hope to obtain new high-speed
photometry to determine once and for all the status of this
object. For the moment, it is the only photometrically constant star
contaminating the strip, but it is entirely possible that we, and Voss
et al., observed HS~1612+5528 during a period of destructive
interference. Alternatively, HS 1612+5528 could also represent a ZZ
Ceti star whose pulsations are not detectable due to our line of sight
with respect to the pulsation modes of the star.

Finally, we see that our survey as once again uncovered several new
objects which lie near or within the empirical boundaries of the strip
and we are exploring the possibility of obtaining high-speed
photometric measurements for these stars as well.

\section{DA+dM binary systems}

Although the majority of stars in our survey are isolated DA white
dwarfs, there are several objects whose optical spectra contain the
tell-tale signs pointing to the presence of a main sequence companion,
usually an M dwarf. The spectrum of the M dwarf will contaminate one
or several of the Balmer lines from the white dwarf spectrum which
renders our normal technique of fitting the observed Balmer line
profiles very difficult. To try and get around this problem, we either
exclude certain spectral lines or certain portions of the spectral
lines from our fitting procedure. However, this means that our
determinations of the atmospheric parameters for these stars are quite
uncertain. In an effort to determine more accurately the atmospheric
parameters of these white dwarfs, we have begun gathering spectra which
cover the entire Balmer series from H$\alpha$ to H8 thus providing
sufficient wavelength coverage to determine the spectral type of the
M dwarf. 

\begin{figure}
\begin{center}
\includegraphics[bb=75 75 550 700,angle=270,width=12cm]{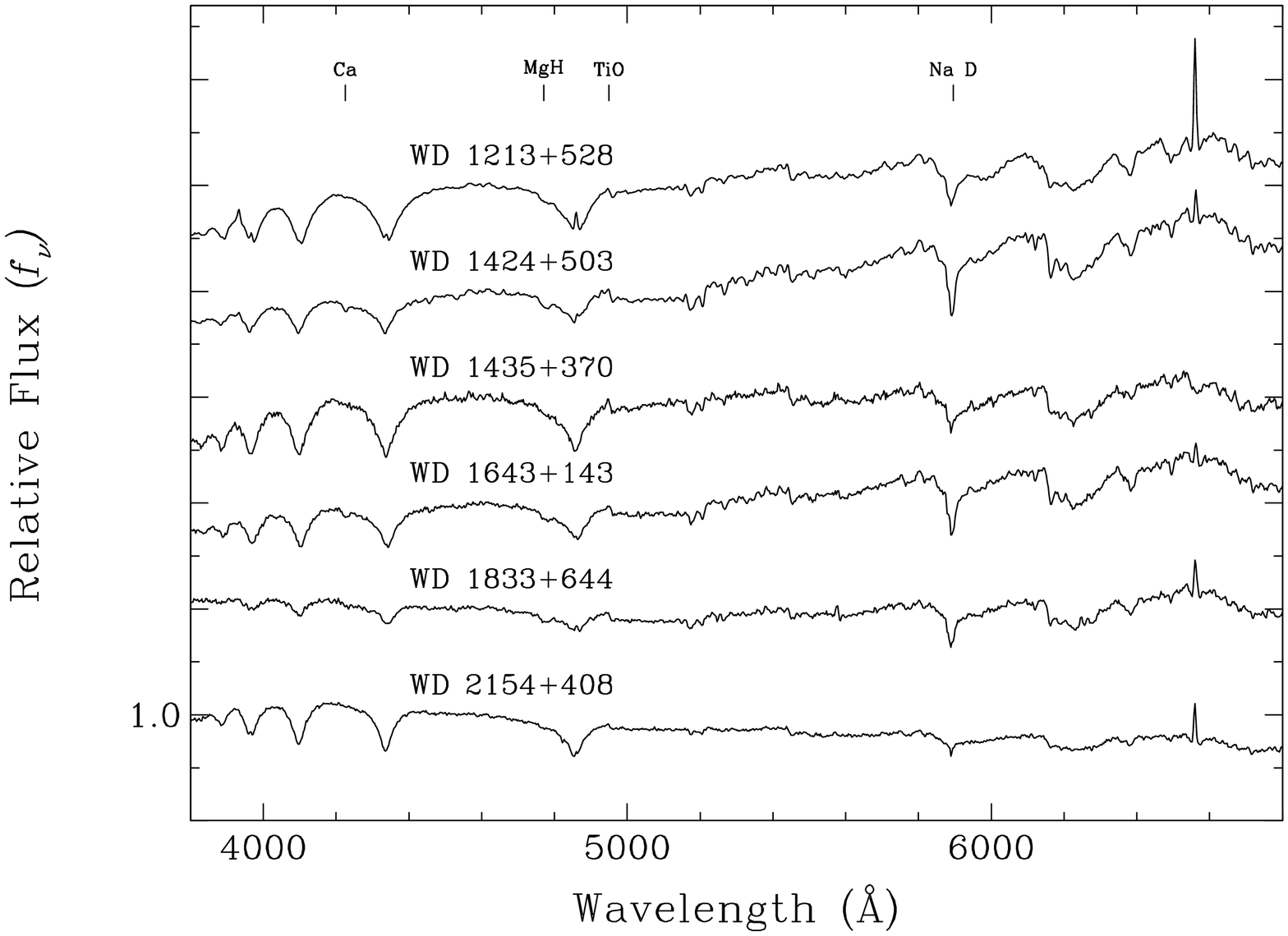}
\caption{\label{DAdM} Optical spectra of DA+dM binary systems. Tick
  marks indicate metallic features in the spectrum of the M dwarf
  companion's spectrum.}
\end{center}
\end{figure}

Our analysis will involve a 5-parameter fit to the entire spectrum:
the effective temperature and surface gravity of the white dwarf
primary, the spectral type of the M dwarf secondary and the relative
intensities of the two energy distributions. We will essentially be
co-adding our usual synthetic spectra of DA white dwarfs with the M
dwarf spectral templates from Bochanski et al.~(2007). We will then use
the results of the fit to subtract an appropriate template, with the
proper flux level, from our initial spectrum. We see in Figure
\ref{DAdM} a sample of some of the spectra we have obtained so
far. The kink we see in the red wing of H$\beta$ is due to a TiO
band. Some spectra also show a MgH band in the blue wing of H$\beta$
as well as Ca {\sc i} between H$\gamma$ and H$\delta$. We also notice
the prominent Na D line at 5895 \AA. Furthermore, hydrogen-line
emission from the M dwarf can contaminate the center of the Balmer
lines of the white dwarf.

\section{CBS 229}

\begin{figure}
\begin{center}
\includegraphics[bb=75 100 550 700,angle=270,width=10cm]{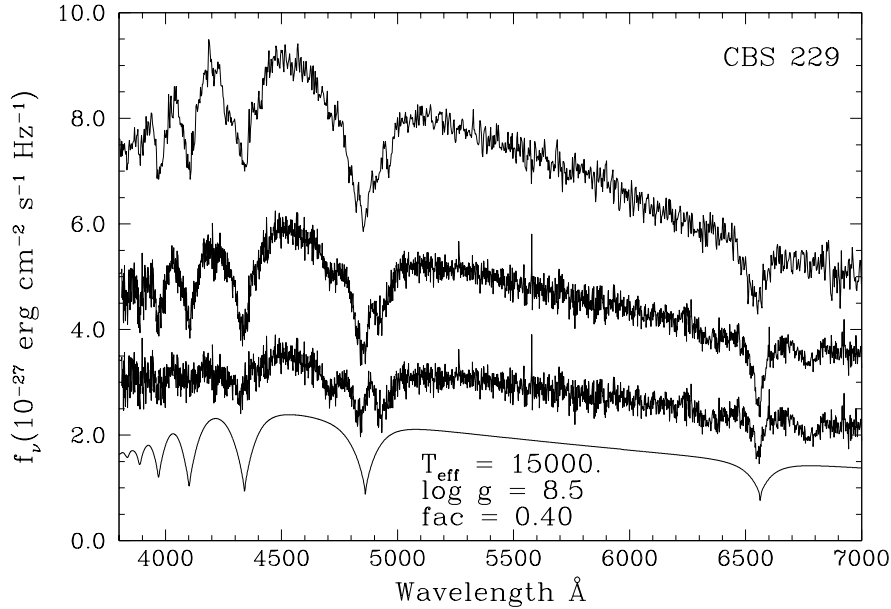}
\caption{\label{CBS} Our own spectrum (first from the top) and the 
SDSS spectrum (second from the top) of CBS 229. At the bottom, a
synthetic spectrum of a DA white dwarf corresponding to the
atmospheric parameters indicated in the figure. The quantity
$fac=0.40$ implies that the synthetic spectrum contributes 40\% of the
flux of the sytem at 5500 \AA. The spectrum just above is the residual
left over from subtracting the synthetic DA model from the SDSS
spectrum.}
\end{center}
\end{figure}

Large surveys often uncover unique and interesting objects.  In the
course of our survey of the DA+dM binaries described in the previous
section, we came across CBS 229, an object which turned out to be
rather interesting. We had initially thought that the feature in the
red wing of H$\beta$ observed in our blue spectrum (not shown here)
was the usual TiO absorption band produced by the presence of an M
dwarf. But our full optical spectrum, shown at the top of Figure
\ref{CBS}, does not show any signs of an M dwarf companion. 
We became aware after the fact that CBS 229 had also been observed as
part of SDSS, and classified as magnetic with an estimated polar
magnetic field of $B_p\sim20$ MG (Vanlandingham et al.~2005). Indeed,
the SDSS spectrum (also displayed in Fig.~\ref{CBS}) with a better
spectral resolution shows what are clearly magnetic components near
H$\alpha$. But the observed profile at H$\beta$, or even at H$\alpha$,
is really at odds with the predictions of magnetic models. For
instance, the models for KPD 0253+5052 shown in Figure 6 of
Wickramasinghe \& Ferrario (2000), with a comparable magnetic field,
predict a much weaker H$\alpha$ central Zeeman component with
respect to the shifted components than observed in CBS
229. Furthermore, the predicted H$\beta$ profile and the higher Balmer
lines are totally smeared out, in sharp contrast with the strong and
sharp Balmer lines observed in CBS 229. 

Instead, we suggest that CBS 229 is an unresolved double degenerate
binary composed of a magnetic DA star and a normal DA star. We show in
Figure \ref{CBS} a very preliminary attempt to deconvolve the spectrum
into its two separate components. We subtracted from the SDSS data a
synthetic spectrum with $T_{\rm eff}=15,000$ K and $\log g=8.5$. We
assumed that the non-magnetic DA contributes 40\% of the flux at 5500
\AA; this sets the relative intensities of the spectra. The residual
spectrum clearly shows magnetic features at H$\alpha$ and H$\beta$
that bear a strong resemblance with the 17 MG model shown in Figure 6
of Wickramasinghe \& Ferrario (2000). To our knowledge, there are not
many double degenerate systems with only one magnetic component that
have been discovered and analyzed, with the exception of the LB
11146 system reported by Liebert et al.~(1993), for which the companion
to the non-magnetic DA white dwarf turned out to be a highly magnetic
helium-atmosphere white dwarf. We plan a more detailed analysis of
CBS 229 that can hopefully shed some more light on the evolution of
such unique systems.

\section{Future Outlook}

We have several upcoming observing runs at Kitt Peak during which we
hope to complete our survey of the McCook \& Sion catalog in the
northern hemisphere. In parallel, we will continue to obtain spectra
for the remaining DA+dM binary systems that we have identified in our
sample. We are also securing telescope time in Chile to complete our
survey in the southern hemisphere. We hope that within a year's time
our survey will essentially be complete.

Our analysis of these stars is ongoing on several fronts. We are
working on adding the necessary physics in our models to analyze the
DAO stars exhibiting the Balmer line problem. Our analysis of the
DA+dM binaries, and in particular, our ability to extract the white
dwarf spectrum will soon yield results as will our analysis of CBS
229.

\ack

We would like to thank the director and staff of Steward Observatory
for the use of their facilities. We would also like to thank
G.~Fontaine for his assistance in conducting the high-speed photometry
and analysis of HS 1612+5528. This work was supported in part by the
NSERC Canada and by the Fund FQRNT (Qu\'ebec). A.~G.~acknowledges the
contribution of the Canadian Graduate Scholarships. P.~B.~is a
Cottrell Scholar of Research Corporation for Science
Advancement. M.~T.~R.~received support from FONDAP 15010003 and PFB-06
(CONICYT).

\section*{References}
\begin{thereferences}
\item Bergeron P, Gianninas A, and Boudreault S 2007 {\it ASP Conf. Series} vol 372, eds R Napiwotzki and M R Burleigh (San Francisco: ASP) p 29  
\item Bergeron, P, Saffer, R and Liebert, J 1992, {\it ApJ} {\bf 394} 247
\item Bochanski J J, West A A, Hawley S L, and Covey K R 2007 {\it AJ} {\bf 133} 531
\item Eisenstein et al. 2006 {\it ApJS} {\bf 167} 40
\item Gianninas A, Bergeron P, Fontaine G 2005 {\it ApJ} {\bf 631} 1100
\item Koester D et al. 2001 {\it A\&A} {\bf 378} 556
\item Liebert J, Bergeron P, and Holberg J B 2005 {\it ApJS} {\bf 156} 47
\item Liebert J, Bergeron P, Schmidt, G D, and Saffer, R A 1993 {\it ApJ} {\bf 418} 426
\item McCook G P and Sion E M 1999 {\it ApJS} {\bf 121} 1
\item Napiwotzki R 1992 {\it Lecture Notes in Physics} vol 401, ed U Heber and C S Jeffery (Berlin: Springer) p 310
\item Napiwotzki R 1999 {\it A\&A} {\bf 350} 101
\item Vanlandingham, K M et al. 2005 {\it ApJ} {\bf 130} 734
\item Voss B, Koester D, Ostensen R, Kepler S O, Napiwotzki R, Homeier D, and Reimers D 2006 {\it A\&A} {\bf 450} 1061
\item Werner K 1999 {\it ApJ} {\bf 457} L39
\item Wickramasinghe D T and Ferrario L 2000 {\it PASP} {\bf 112} 873
\end{thereferences}
\end{document}